\documentclass[prb,twocolumn,showpacs,preprintnumbers,amsmath,amssymb,superscriptaddress]{revtex4}

\usepackage{threeparttable}%
\usepackage{color}%
\usepackage{graphicx}
\usepackage{dcolumn}
\usepackage{bm}
\usepackage{float}

\begin{document}


\title{Synthesis and characterization of multiferroic BiMn$_7$O$_{12}$}

\author{F.~Mezzadri}
\email[Corresponding Author.~E-mail: ]{francesco.mezzadri@nemo.unipr.it}
\affiliation{Dipartimento di Chimica GIAF, Universit$\grave{a}$ di Parma, Viale G.P. Usberti 17A, I-43100 Parma, Italy}

\author{G.~Calestani}
\affiliation{Dipartimento di Chimica GIAF, Universit$\grave{a}$ di Parma, Viale G.P. Usberti 17A, I-43100 Parma, Italy}
\affiliation{Istituto dei Materiali per Elettronica e Magnetismo IMEM-CNR,
Area delle Scienze, 43100 Parma, Italy}
\author{M.~Calicchio}
\author{E.~Gilioli}
\author{F.~Bolzoni}
\author{R.~Cabassi}
\affiliation{Istituto dei Materiali per Elettronica e Magnetismo IMEM-CNR,
Area delle Scienze, 43100 Parma, Italy}
\author{M.~Marezio}
\affiliation{CRETA CNRS, 38042 Grenoble Cedex 9, France}
\author{A. Migliori}
\affiliation{CNR-IMM, Via Gobetti 101, I-40126 Bologna, Italy}

\begin{abstract}
We report on the high pressure synthesis of BiMn$_7$O$_{12}$, a manganite displaying a "quadruple perovskite"
structure. Structural characterization of single crystal samples shows a distorted and asymmetrical
coordination around the Bi atom, due to presence of the $6s^{2}$ lone pair, resulting in non-centrosymmetric
space group Im, leading to a permanent electrical dipole moment and ferroelectric properties. On the
other hand, magnetic characterization reveals antiferromagnetic transitions, in agreement with the isostructural
 compounds, thus evidencing two intrinsic properties that make BiMn$_7$O$_{12}$ a promising multiferroic
  material.
\\
\end{abstract}

\maketitle

Multiferroics are defined as materials that simultaneously exhibit more than one ferroic order
parameter among ferromagnetism, ferroelectricity and ferroelasticity.~\cite{MULTI} Although there are a lot of compounds
 presenting magnetic or ferroelectric order, the constrains required for their coexistence are so severe
 that only an extremely limited number of multiferroic materials exists. Besides scientific interest
 in their physical properties, multiferroics have potential for technological applications as actuators,
  switches, magnetic field sensors or new types of electronic memory devices. However, at present, only magnetoelectric
  composites, realized by combining magnetostrictive and piezoelectric materials, are ready for technological
  applications.~\cite{COMP} Typical multiferroics belong to the group of the perovskitic transition metal oxides,
  and include rare-earth manganites and ferrites
  (TbMnO$_{3}$,~\cite{TBMNO} HoMn$_{2}$O$_{5}$,~\cite{HOMNO} LuFe$_{2}$O$_{4}$~\cite{LUFEO}), bismuth-based compound
  BiFeO$_{3}$,~\cite{BIFEO} non-oxides (BaNiF$_{4}$~\cite{BANIF}) and spinel chalcogenides (ZnCr$_{2}$Se$_{4}$~\cite{ZCS}).
 In usual perovskite-based materials, the ferroelectric distortion occurs due to the displacement of B-site
  cation (for example, Ti in BaTiO$_{3}$~\cite{BTO}) with respect to the oxygen octahedral coordination. One possible
   mechanism for the coexistence of ferroelectricity and magnetism,
\begin{figure}
\includegraphics[width=9.0cm]{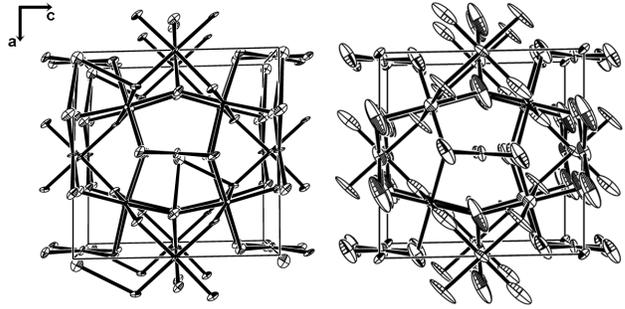}
\caption{\label{fig:B}ORTEP perspective view of the BiMn$_7$O$_{12}$ structure refined in Im (left)
and I2/m (right) space groups. Displacement ellipsoids are drawn in both cases at the 50\%
probability level.}
\end{figure}
is the presence in the A-site of an atom carrying a non-bonding pair of electrons in an outer shell
   (lone-pair), whose stereochemical effect may be at the origin of ferroelectricity, while partially filled d orbitals
   on the B site should be responsible for the magnetic behavior. Examples include BiFeO$_{3}$~\cite{BIFEO} and PbVO$_{3}$.~\cite{PBVO}
    In these materials, the $6s^{2}$ lone-pair on the A-site cation (Bi$^{3+}$, Pb$^{2+}$) causes the Bi/Pb $6p$ (empty) orbital to come
   closer in energy to the O $2p$ orbitals. This leads to the hybridization between the Bi/Pb $6p$ and O $2p$ orbitals and drives
   the off-centering of the cation towards the neighboring anion resulting in ferroelectricity.\\
 BiMn$_7$O$_1$$_2$ is a metastable material, belonging to the family of "quadruple-perovskite"\footnotemark[1] \  manganites, with
 general formula AA'$_{3}$B$_{4}$O$_{12}$,~\cite{NAMNO,CAMNO,LAMNO} derived by the doubling of the
 conventional ABO$_{3}$ manganites axes.
  \footnotetext[1]{"quadruple" refers to the formula unit ABO$_3$}
  \begin{table*}
\caption{\label{tab:tableA}Atomic coordinates and parameters for BiMn$_7$O$_1$$_2$. U$_e$$_q$ is
defined as one third of the trace of the orthogonalized U$_i$$_j$ tensor. Anisotropic displacement
parameters:  U$_i$$_j$ = exp(-2$\pi$$^2$(U$_1$$_1$h$^2$(a*)$^2$ + ... + 2U$_1$$_2$hk(a*)(b*) + ...).}
\begin{ruledtabular}
\begin{tabular}{lcccccccccc}
 Atom&x&y&z&U$_e$$_q$&U$_1$$_1$&U$_2$$_2$&U$_3$$_3$&U$_2$$_3$&U$_1$$_3$&U$_1$$_2$\\
\hline
Bi1&0.0284(6)&0&0.0173(6)&0.0255(3)&0.0247(8)&0.0307(3)&0.0210(5)&0&-0.0028(4)&0 \\
Mn1&0.0030(30)&0&0.4951(3)&0.0106(6)&0.0066(15)&0.0185(10)&0.0066(16)&0&-0.0041(10)&0\\
Mn2&0.4997(4)&0&0.5019(4)&0.0102(6)&0.014(2)&0.0101(9)&0.0069(13)&0&-0.0041(11)&0\\
Mn3&0.4955(2)&0&-0.0022(4)&0.0101(6)&0.0034(14)&0.0082(9)&0.0185(14)&0&-0.0049(10)&0\\
Mn4&0.2542(3)&0.7419(3)&0.2540(3)&0.0083(4)&0.0080(10)&0.0076(6)&0.0093(8)&-0.0020(7)&-0.0031(7)& -0.0030(8)\\
Mn5&0.2476(2)&0.74930(17)&0.74590(18)&0.0067(5)&0.0066(12)&0.0084(6)&0.0049(10)&-0.0016(5)& -0.0046(7)& 0.0006(6)\\
O1&0.3366(19)&$\frac{1}{2}$&0.1827(18)&0.015(2)&0.006(5)&0.023(5)&0.015(6)&0&-0.007(4)&0\\
O2&0.1916(19)&0&0.6755(19)&0.017(2)&0.010(6)&0.023(5)&0.017(6)&0&-0.007(4)&0\\
O3&0.1742(18)&0&0.3031(18)&0.017(2)&0.004(5)&0.029(6)&0.017(6)&0&-0.001(4)&0\\
O4&0.825(2)&0&0.3094(19)&0.020(3)&0.012(6)&0.029(6)&0.017(6)&0&-0.010(5)&0\\
O5&0.4910(16)&0.8086(13)&0.3282(14)&0.0196(19)&0.016(4)&0.028(4)&0.015(4)&-0.015(3)&-0.006(3)& -0.006(4)\\
O6&0.3081(15)&0.8259(10)&-0.0105(14)&0.0192(15)&0.019(4)&0.016(4)&0.023(4)&0.001(3)&-0.009(3)& -0.001(3)\\
O7&0.0167(14)&0.6830(13)&0.1765(15)&0.0197(19)&0.009(4)&0.031(4)&0.019(4)&-0.010(3)&-0.006(3)& -0.010(4)\\
O8&0.6796(15)&0.1800(10)&0.0112(13)&0.0162(15)&0.017(4)&0.015(4)&0.016(4)&0.000(3)& -0.008(3)& -0.002(3)\\
\end{tabular}
\end{ruledtabular}
\end{table*}
\begin{table}[]
\caption{\label{tab:tableB} Interatomic distances (\AA).}
\begin{ruledtabular}
\begin{tabular}{ccccc}
 Atoms&Distance&&Atoms&Distance\\
\hline
Bi1-O3&2.392(14)&&Mn3-O6&2x1.910(10)\\
Bi1-O6&2x2.481(9)&&Mn3-O8&2x1.923(10)\\
Bi1-O7&2x2.631(10)&&&\\	
Bi1-O5&2x2.697(10)&&Mn4-O7&1.920(12)\\	
Bi1-O4&2.706(13)&&Mn4-O5&1.922(13)\\	
Bi1-O2&2.872(15)&&Mn4-O1&1.969(6)\\	
Bi1-O1&2.874(14)&&Mn4-O3&2.035(5)\\
Bi1-O8&2x2.945(11)&&Mn4-O8&2.076(11)\\
                 &&&Mn4-O6&2.130(11)\\	
Mn1-O1&1.907(14)&&&\\			
Mn1-O4&1.916(16)&&Mn5-O8&1.897(10)\\
Mn1-O2&1.943(16)&&Mn5-O6&1.961(11)\\
Mn1-O3&1.956(13)&&Mn5-O2&1.969(6)\\
			   &&&Mn5-O4&1.986(6)\\	
Mn2-O7&2x1.887(11)&&Mn5-O5&2.087(12)\\	
Mn2-O5&2x1.925(10)&&Mn5-O7&2.163(11)\\
\end{tabular}
\end{ruledtabular}
\end{table}
The complex and highly distorted structure is based on a 3D network of corner-sharing
  MnO$_{6}$ tilted octahedra, centred on the B site; it can only be accommodated under high pressure by
  the presence of a Jahn-Teller atom (Mn$^{3+}$ or Cu$^{2+}$) on the A' site, displaying an uncommon square
   planar coordination due to large distortion of the standard dodecahedral coordination, typical of the A site.
    Since the occupation of the A site determines the properties of the compounds, the motivation of this work
    is the synthesis of the Bi-substitute member and to correlate the structural distortion induced by the Bi$^{3+}$
    lone pair to the electronic properties, in particular searching for multiferroic properties.\\
   BiMn$_7$O$_{12}$ was synthesized by solid state reaction in high pressure/high temperature (HP/HT) conditions.
   Stoichiometric mixture of Mn$_2$O$_3$ (Ventron 98\%) and Bi$_2$O$_3$ (Merck 99\%) was used as reagent.
   They were mixed, finely grounded in glove box and encapsulated in a Pt foil inserted in the MgO octahedral cell and
   in the multi anvil apparatus. The pressure was increased up to its maximum value at a rate of 160 bar/min,
   then the capsule was heated up to the reaction temperature at a rate of 50$^{\circ}$C/min. The optimal conditions
    for the synthesis of a mixture of bulk material and single crystals were identified as 40 Kbar and 1000$^{\circ}$C.
    After 2 hours in these conditions the sample was cooled down to room temperature by switching off the heater.
    The pressure was finally slowly released at 0.4 bar/min.
\begin{figure*}[]
\includegraphics[width=11cm,viewport=0 0 600 450,clip]{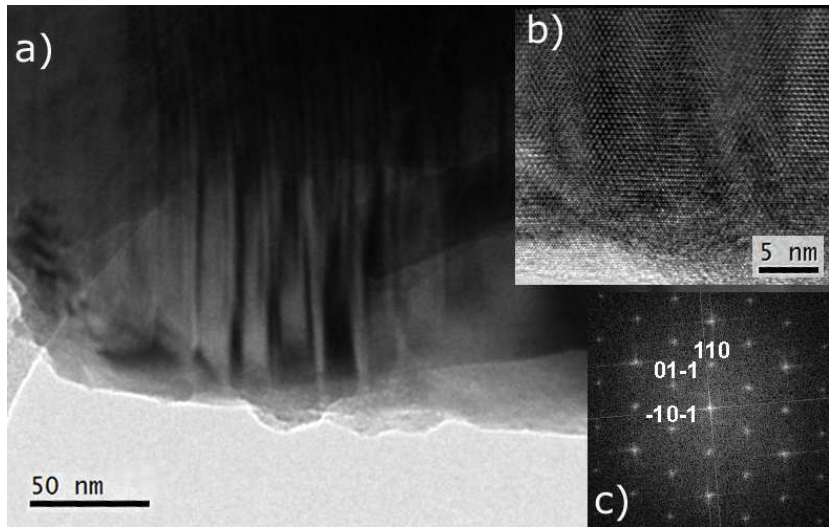}
\caption{\label{fig:C}Bright Field (a) and High Resolution (b) TEM images taken in the [-111]
zone axis showing the presence of twinning domains in a BiMn$_7$O$_1$$_2$ sample.
Fast fourier transform of the HREM image is shown in (c).}
\end{figure*}

The crystal structure of BiMn$_7$O$_1$$_2$ at RT was determined by single crystal X-ray diffraction.
Intensity data were collected by using MoK$_a$ data in the range 3.79 $\le$  $\theta$ $\le$ 28.83$^{\circ}$ on
a Bruker AXS Smart diffractometer, equipped with a CCD area-detector. The structure is monoclinic, with lattice
parameters \emph{a} = 7.5351(15), \emph{b} = 7.3840(15), \emph{c} =  7.5178(15) {\AA} and $\beta$ = 91.225(3)$^{\circ}$.
The cell volume is V = 418.19(15) \AA$^3$, consistent with 2 formula units
per cell. In analogy with other monoclinic AMn$_7$O$_1$$_2$ compounds, the structure was initially refined
with SHELX97~\cite{SHELX} in the  centrosymmetric space group I2/m, using anisotropic atomic displacements
parameters (a.d.p.'s) for all the atoms. The refinement converged to R$_1$ = 0.0535, wR$_2$ = 0.1413
and g.o.f = 1.101 for 562 data and 59 parameters. In spite
  of the fact that these final agreement indices can be considered satisfactory, when the sole numeric
   values are considered, a careful inspection of the refined parameters points out a very unusual trend of
   the a.d.p.'s showing, in particular for the oxygen atoms, not only very high values but also an anomalous
 elongation in a common direction of the \emph{ac} plane  (Fig.\ref{fig:B}), that indicates static disorder.
Therefore the possibility of an artifact induced by forcing an acentric structure in a centric symmetry was
 taken into account and the structure was further refined in the noncentrosymmetric I2 and Im space groups.
  While the results obtained in I2 are
comparable to those of the centrosymmetric model, the ones produced in  Im are characterized
by a significant improvement in terms of agreement indices (R$_1$ = 0.0391, wR$_2$ = 0.0928 and
g.o.f = 1.113 for 1082 data and 103 parameters) and a.d.p.'s. The latter is clearly evidenced
in Fig.1, that shows a comparison of the ORTEP~\cite{ORTEP} plots of the crystal structure of BiMn$_7$O$_1$$_2$
refined in both Im and I2/m, in which the a.d.p.'s are represented by ellipsoids drawn at the 50\% probability
level. Atomic parameters, refined in the acentric space group Im, and relevant bond distances are shown
in Table \ref{tab:tableA} and \ref{tab:tableB}, respectively. Charge
 distribution analysis, performed with CHARDIS,~\cite{CHARDIS} indicates a 3+ oxidation state for all the Mn atoms.
The decrease of symmetry from the centrosymmetric space group I2/m, typical of other monoclinic
  A$^{3+}$Mn$_7$O$_1$$_2$ compounds, to Im in BiMn$_7$O$_1$$_2$ is mainly determined by the steric hindrance
  of the 6s$^2$ lone pair of the Bi$^{3+}$ ion. This typically produces a
quite distorted coordination
   around the Bi atom, with the strongest bonds lying on the same side in order to accommodate the lone
\begin{figure}
\centering
\includegraphics[width=8.5cm,viewport=0 0 500 600,clip]{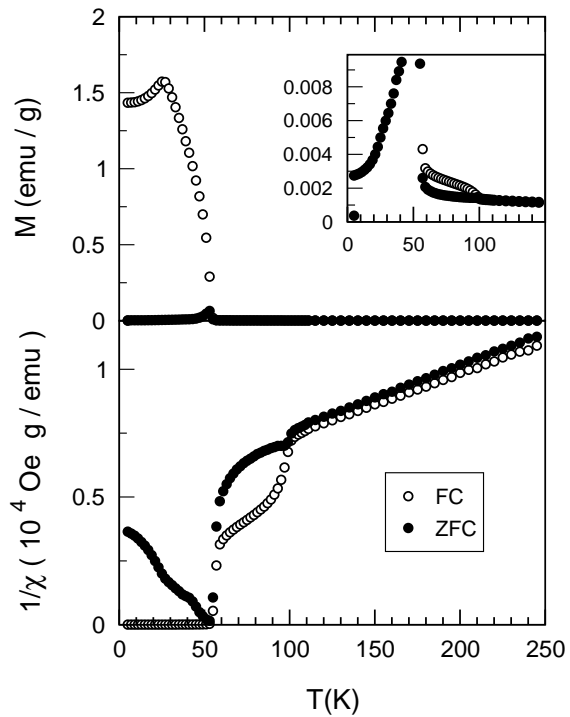}
\caption{\label{fig:A}Zero Field Cooled (ZFC, filled symbols) and Field
Cooled (FC, open symbols) DC magnetization curves (upper panel) and
inverse susceptibility (lower panel) in applied magnetic field H = 10
Oe. Inset: secondary  transition in the magnetization curve  at T
$\approx$ 100K.}
\end{figure}
   pair on the opposite one. Structural data suggest that the Bi atom moves out from the center of the
  icosahedral oxygen coordination determining an asymmetrical coordination.
An electrical polarization of 7.33 $\mu$C/cm$^2$ at RT, generated by the displacement of all the
Bi atoms in a common direction, can be computed by a simple model as dipole moment
for volume unit, suggesting the existence of ferroelectricity in BiMn$_7$O$_{12}$.
A similar value was reported~\cite{BIFEO} for the simple perowskite BiFeO$_3$, where a spontaneous polarization
of 6.1 $\mu$C/cm$^2$ was measured at 77 K on bulk samples.
This hypothesis is further supported by TEM investigations, performed on a
Philips TECNAI F20 transmission electron microscope operating at 200 KV,
which show the presence of twinning domains extending on a few tens of
nanometers, evidenced in Fig.\ref{fig:C}a and Fig.\ref{fig:C}b by sharp contrast variations. The
domains orientation is consistent with the inversion of the polar axis,
namely the shift of the Bi ions in opposite directions.\\
The magnetic properties of BiMn$_7$O$_1$$_2$ have been studied using a
SQUID magnetometer. The measurements were carried out on polycrystalline
samples because the single crystals yielded a too weak signal, approaching
the instrumental sensitivity. The Zero Field Cooled (ZFC) and Field Cooled (FC) DC magnetization curves in applied
magnetic field H = 10 Oe are reported in Fig.\ref{fig:A} (upper panel), where a sharp
transition to antiferromagnetic regime at T$_c$ $\approx$ 50 K can be seen. The antiferromagnetic nature of
the ordered phase is more clear looking at the inverse susceptibility shown in Fig.\ref{fig:A}
(lower panel), where one can also estimate the Curie-Weiss temperature $\theta$ $\approx$ 206 K.
The anomaly at T $\approx$ 100 K could be ascribed to the possible presence of  BiMnO$_3$
impurities, whose monoclinic phase undergoes a ferromagnetic transition at T$_c$ = 99 K~\cite{A}
recoverable in an amplified view of the magnetization curve (inset of Fig.\ref{fig:A}).
 From the magnetization of BiMnO$_3$~\cite{B} one can assess that the impurity would be less
 than 1\%. The sharp drop of the inverse susceptibility at T$_c$ has the shape usually observed in
 stoichiometrically pure Dzyaloshinskii-Moriya compounds,\cite{C,D} which is consistent with the MnO$_6$
octahedra tilting. It is worth to remark that a similar behavior is also observed in the structurally
  related compounds PrMn$_7$O$_1$$_2$~\cite{PRMNO} and LaMn$_7$O$_1$$_2$,~\cite{LAMNO} while is absent in the
   simple perovskite BiMnO$_3$. A detailed study of the magnetic structure by neutron diffraction is in
   progress. The slope of the inverse susceptibility yields $\approx$ 6 $\mu_B$/Mn, while the expected value
   for Mn$^{3+}$ ions would be 4.9 $\mu_B$/Mn. The antiferromagnetic peak at 26 K should correspond to the ordering
    of the Mn ions in the planarly coordinated  A' sites.\\
In conclusion, we have synthesized by HP/HT BiMn$_{7}$O$_{12}$, a new member of the intriguing family of
manganites with quadruple perovskite structure. Detailed structural characterization carried out by single
crystal X-ray diffraction yields the definition of a non-centrosymmetric space group (Im) never observed
before in this class of compounds. This feature is originated by the stereochemical effect induced by the presence
of the $6s^{2}$ lone pair of the Bi$^{3+}$ ions, that induces an asymmetrical coordination of the oxygen
neighbors, leading to a permanent electrical dipole moment.
 Magnetic characterization points out, in agreement with isostructural compounds, the
 presence of antiferromagnetic transitions at 50 and 26 K should correspond to the ordering of the B and A' sites
 respectively. The coexistence of the two ferroic orders makes of BiMn$_{7}$O$_{12}$ a new promising
 multiferroic material.\\
The authors wish to express their gratitude to G. Andr\'{e}, F. Bour\'{e}e, A. Gauzzi, G. Rousse, A. Prodi, F. Licci for
fruitful discussions and to T. Besagni for XRD.
A few high pressure experiments of BiMn$_7$O$_{12}$, were performed at the Bayerisches Geoinstitut at Bayreuth
 (Germany) under the EU "Research Infrastructures: Transnational Access" Programme (Contract No.505320 (RITA)-
 High Pressure).\\

\end{document}